\documentclass[aps,prb,twocolumn,floatfix,showpacs,citeautoscript,superscriptaddress,longbibliography,hyperlinks]{revtex4-2}
\usepackage{amsmath}
\usepackage{graphicx}
\usepackage{epstopdf}

\usepackage[colorlinks=true,citecolor=blue]{hyperref}
\usepackage{float}
\usepackage[caption = false]{subfig}

\newcommand{\cumu}[1]{\langle \! \langle #1 \rangle \! \rangle}

\begin{document}
\title{Nonequilibrium phase transition in a single-electron micromaser}
\author{Fredrik Brange}
\affiliation{Department of Applied Physics, Aalto University, 00076 Aalto, Finland}
\author{Aydin Deger}
\affiliation{Department of Applied Physics, Aalto University, 00076 Aalto, Finland}
\affiliation{Interdisciplinary Centre for Mathematical Modelling and Department of Mathematical Sciences,
  Loughborough University, Loughborough, Leicestershire LE11 3TU,
  United Kingdom}
\author{Christian Flindt}
\affiliation{Department of Applied Physics, Aalto University, 00076 Aalto, Finland}

\begin{abstract}
Phase transitions occur in a wide range of physical systems and are characterized by the abrupt change of a physical observable in response to the variation of an external control parameter. Phase transitions are not restricted to equilibrium situations but can also be found in nonequilibrium settings, both for classical and quantum mechanical systems. Here, we investigate a nonequilibrium phase transition in a single-electron micromaser consisting of a microwave cavity that is driven by the electron transport in a double quantum dot. For weak electron-photon couplings, only a tiny fraction of the transferred electrons are accompanied by the emission of photons into the cavity, which essentially remains empty. However, as the coupling is increased, many photons are suddenly emitted into the cavity. Employing ideas and concepts from full counting statistics and Lee-Yang theory, we analyze this nonequilibrium phase transition based on the dynamical zeros of the factorial moment generating function of the electronic charge transport statistics, and we find that the phase transition can be predicted from short-time measurements of the higher-order factorial cumulants. These results pave the way for further investigations of critical behavior in open quantum systems.
\end{abstract}

\maketitle

\section{Introduction}

Micromasers allow for detailed investigations of light-matter interactions in quantum cavity electrodynamics~\cite{Englert,MeystreSargent2007,PhysRevA.69.042302}. A micromaser consists of a high-quality photon cavity, which is driven by excited two-level systems, such as a beam of atoms~\cite{PhysRevLett.104.160601,PhysRevE.84.021115}, a superconducting single-electron transistor~\cite{PhysRevLett.98.067204,PhysRevE.85.051122,Armour:2017,Kubala:2020}, or a double quantum dot~\cite{PhysRevA.69.042302,Bergenfeldt:2013,Bergenfeldt:2014,Liu285,Mantovani:2019,Berg:2019}. Several experiments~\cite{PhysRevLett.54.551,PhysRevLett.58.353,PhysRevLett.73.3375,PhysRevA.56.1662,Nogues1999,PhysRevLett.83.5166,Rauschenbeutel2000,PhysRevLett.86.3534,Vahala2003,McKeever2003,PhysRevLett.103.167402,PhysRevLett.113.036801,Menard2022} with micromasers and their optical counterparts, microlasers, have revealed a variety of intriguing quantum phenomena such as quantum collapse and revival~\cite{PhysRevLett.58.353}, the generation of nonclassical atom statistics~\cite{PhysRevLett.83.5166,PhysRevLett.86.3534,PhysRevLett.103.167402}, and nondestructive single-photon detection~\cite{Nogues1999}. Micromasers also play an important role in many quantum technologies, for example for realizing quantum heat engines~\cite{Scovil:1959,Menczel:2021}.

\begin{figure}[b!]
    \centering
    \includegraphics[width=0.95\columnwidth]{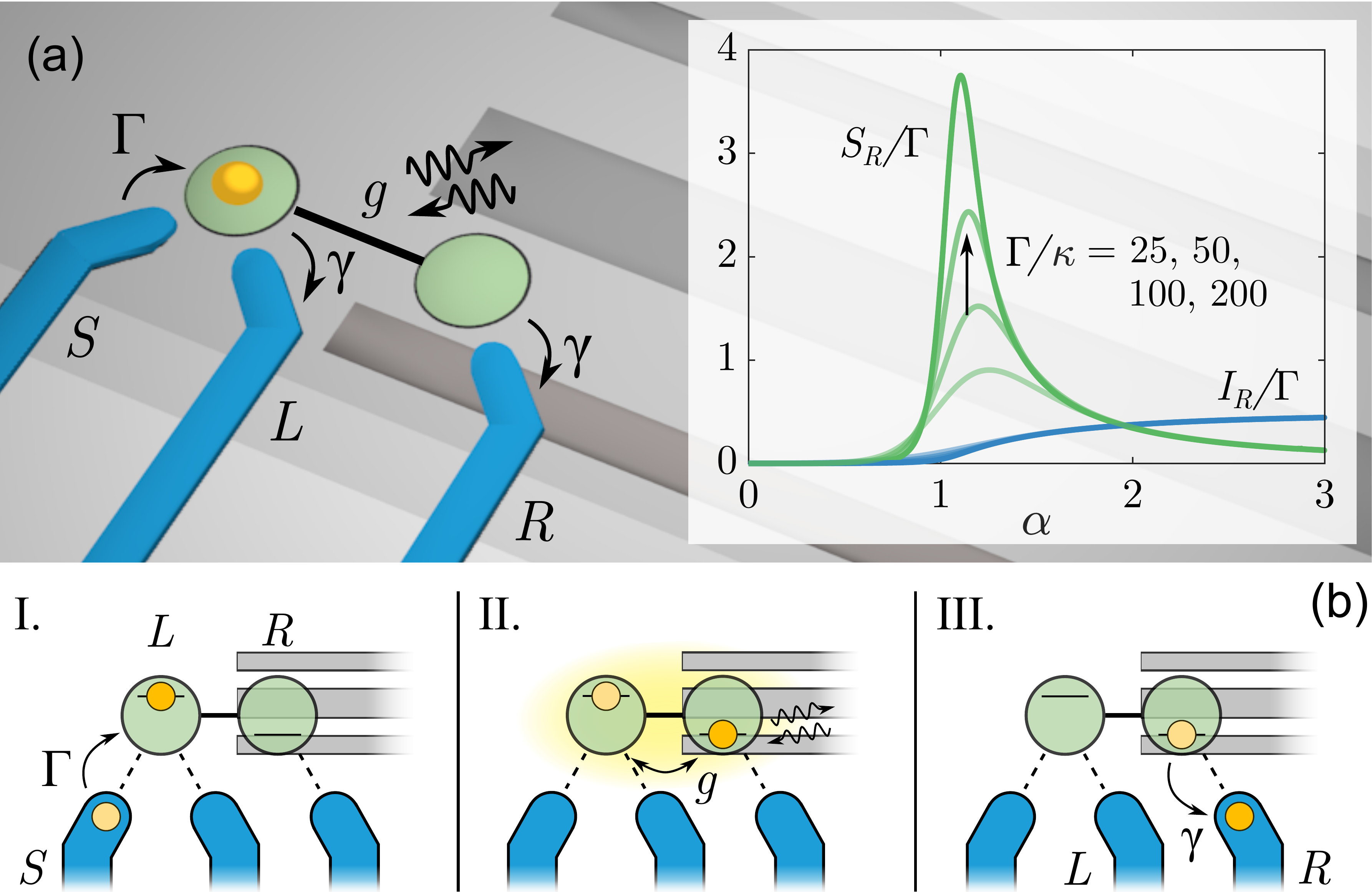}
    \captionsetup{justification=justified,singlelinecheck=false}
    \caption{Single-electron micromaser. (a)~The single-electron micromaser consists of a microwave cavity (dark gray) that is coupled with amplitude $g$ to two quantum dots (light green) and with rate $\kappa$ to a thermal bath (not shown). The average occupation number of the bath is $\nu$. The micromaser is pumped by electrons entering the quantum dots from a biased source electrode (blue, $S$) and leaving through any of the two grounded drains (blue, $L$ and $R$). (b)~Illustration of the single-electron pumping cycle: I.~An electron tunnels into the left dot from $S$ at the rate $\Gamma$. II.~The electron interacts coherently with the cavity, emitting (absorbing) a photon as it tunnels to the right (left) dot. III.~The electron leaves the double dot system through one of the drains (here, $R$) at the rate $\gamma \gg \Gamma$ before the next electron enters from the source. The inset of panel (a) shows the average output current $I_R$ in drain $R$ and the noise $S_R$ (defined in the main text) as functions of the pump parameter $\alpha = g/\gamma\sqrt{2\Gamma/\kappa}$ for different values of $\Gamma$, with $\nu=0.1$. Around $\alpha \simeq 1$, a phase transition from a low-activity to a high-activity phase occurs for $\Gamma\gg\kappa$.}
    \label{Micromaser}
\end{figure}

In a micromaser, the excited two-level systems amplify the photon field by transferring their excitation energy to the cavity~\cite{Englert,MeystreSargent2007,PhysRevA.69.042302}, which may prompt a drastic change in the system dynamics. For weak electron-photon couplings, the cavity essentially remains empty. However, as the coupling is increased, the system may undergo a  phase transition as many photons are suddenly emitted into the cavity. This nonequilibrium phase transition takes place between two distinct dynamical phases, which are characterized by different quantum jump trajectories~\cite{PhysRevLett.104.160601,PhysRevE.84.021115}. Each phase can be described by its full counting statistics of quantum jumps, and in the thermodynamic limit of large system sizes and long observation times, the quantum jump trajectories become very long.

To understand such nonequilibrium phase transitions, the Lee-Yang theory of equilibrium phase transition~\cite{Yang:1952,Lee:1952,Blythe2003,Bena:2005,Biskup2000,PhysRevLett.109.185701,PhysRevLett.114.010601} has been adapted to systems out of equilibrium~\cite{Arndt2000,Blythe2002,Dammer2002,Flindt2013,Hickey2013,Hickey2014}. In the approach that we follow here, one considers the zeros of the moment generating function of the full counting statistics in the complex plane of the counting field~\cite{Flindt2013,Hickey2013,Hickey2014,Deger:2018,Deger:2019,Deger:2020,Deger:2020b,Peotta2021,Kist2021}. The full counting statistics may display signatures of critical behavior~\cite{PhysRevB.81.045317,Padurariu:2012,Arndt:2019,Arndt:2021}, and in case of a phase transition, the zeros should reach the origin of the complex plane in the thermodynamic limit. Moreover, the zeros can be determined from measurements of the higher-order cumulants of the full counting statistics~ \cite{Flindt2013,Hickey2013,Hickey2014,Deger:2018,Deger:2019,Deger:2020,Deger:2020b,Peotta2021,Kist2021}. This idea was realized in the experiment of Ref.~\onlinecite{Brandner2017}, which measured the full counting statistics of electron tunneling in a metallic island to determine the dynamical Lee-Yang zeros. However, the system only involved a few electronic states, such that the thermodynamic limit of large system sizes could not be reached, and no phase transition was observed. In this context, it has recently been realized that the large state space of a single quantum harmonic oscillator is sufficient to reach the thermodynamic limit, and a quantum phase transition has been predicted for the Rabi model~\cite{Hwang2015,Hwang2018}, which involves just a single two-level system coupled to a quantum harmonic oscillator.  

With this in mind, we here consider electron transport through a double quantum dot coupled to a microwave cavity, making up a single-electron micromaser. As we will see, the micromaser exhibits a nonequilibrium phase transition at a critical value of the electron-photon coupling, and the phase transition can be determined from short-time measurements of the full counting statistics as in the experiment of Ref.~\onlinecite{Brandner2017}. We note that similar phase transitions have been predicted for atomic micromasers~\cite{PhysRevLett.104.160601,PhysRevE.84.021115} by treating the counting field as an external control parameter that can be varied at will~\cite{Merolle2005,Garrahan2007,Hedges2009,Speck2012}. Here, by contrast, we show how the phase transitions can be directly observed from accurate measurements of the full counting statistics. In particular, we will see that factorial moments and factorial cumulants~\cite{PhysRevB.83.075432,Kambly2013,Stegmann2015,Stegmann2016,PhysRevB.94.125433,Stegmann2017,Kleinherbers2018,10.21468/SciPostPhys.10.1.007,PhysRevB.104.125431,Kleinherbers2021,besson2021,stegmann2021} are particularly useful to reveal the nonequilibrium phase transition, which may be observed in future experiments.

The paper is organized as follows. In Sec.~\ref{sec:model}, we describe our model of a single-electron micromaser, whose dynamics is governed by a Lindblad master equation. In Sec.~\ref{sec:fcs}, we introduce the full counting statistics of electron tunneling together with the factorial cumulants that we use to characterize the dynamical phases of the micromaser. In Sec.~\ref{Dynamical phases}, we identify the low and high activity phases of the system for weak and strong couplings, and we show that both dynamical phases are characterized by Poissonian full counting statistics. In Sec.~\ref{Non-equilibrium phase transition}, we consider the micromaser as we tune the coupling from weak to strong and observe a nonequilibrium phase transition between the two dynamical phases. To this end, we make use of the Lee-Yang theory of equilibrium phase transitions, adapted to the nonequilibrium situation that we consider here. In Sec.~\ref{sec:expper}, we briefly comment on the perspectives of observing our predictions in future experiments before finally presenting our conclusions and an outlook in Sec.~\ref{sec:conclusions}. Technical details of our calculations are presented in Apps.~\ref{Appendix A} and~\ref{Appendix B}.

\section{Single-electron micromaser}
\label{sec:model}

\begin{figure*}[t]
    \centering
    \includegraphics[width=0.95\textwidth]{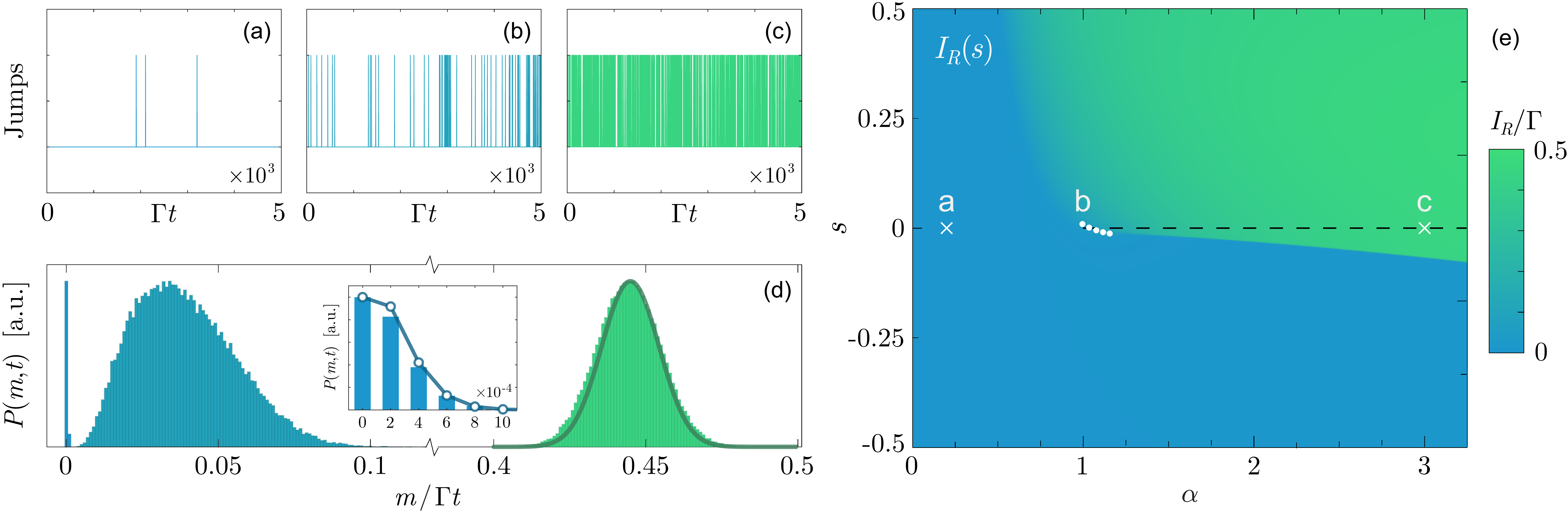}
    \captionsetup{justification=justified,singlelinecheck=false}
    \caption{Simulated time traces of the right drain current for (a) $\alpha=0.2$, (b) $\alpha = 1$ and (c) $\alpha = 3$. (d) The corresponding (unnormalized) probability distributions $P(m,t)$ obtained from $10^5$ simulations for each value of $\alpha$ and $\Gamma t = 5\times 10^3$. For $\alpha=3$, the Gaussian distribution (dark green, solid) from Eq.~\eqref{High activity prob} is plotted for comparison. The inset shows $P(m,t)$ for $\alpha = 0.2$ together with the  Poissonian distribution (dark blue, solid) obtained from Eq.~\eqref{Low activity prob}. (e) Average drain current $I_R(s) = \langle m\rangle_F(s)/t$ as a function of the counting variable $s$ and the pump parameter $\alpha$ obtained by exact diagonalization of the Liouvillian with $N=175$ cavity states. The white dots show the critical value of $s$ for different values around $\alpha = 1$  determined with the cumulant method. The dashed lines shows the phase transition line for $\Gamma\rightarrow\infty$. In all panels, $\nu = 0.1$ and $\Gamma/\kappa=250$. }
    \label{Phase diagrams}
\end{figure*}

The single-electron micromaser in Fig.~\ref{Micromaser}(a) consists of a microwave cavity with resonance frequency $\omega_0$ and decay rate $\kappa$ that is coupled to the electronic transport in a double quantum dot system. Due to strong Coulomb interactions, the quantum dots can in total not  be occupied by more than a single electron at a time. As such, the quantum dots act as a two-level system, which we tune into resonance with the cavity. Specifically, the quantum dot levels are detuned by the energy $\hbar \omega_0$, which is much larger than the tunnel coupling between them, so that the level of the left dot is the excited state, and the right one is the ground state. Similar to how a single-atom micromaser is driven by excited two-level atoms~\cite{Englert,PhysRevE.84.021115}, the single-electron micromaser is pumped by individual electrons passing through the quantum dots as illustrated in Fig.~\ref{Micromaser}(b). Electrons tunnel into the left quantum dot from a biased source electrode at the rate $\Gamma$. Once inside the quantum dots, electrons interact coherently with the photon field of the cavity with the interaction strength~$g$. A photon may thereby be emitted into the cavity as an electron is transferred from the left to the right quantum dot, or, similarly, a photon can be absorbed from the cavity as the electron is transferred from the right to the left quantum dot. These processes continue until the electron eventually leaves the quantum dots via the left or right drains at the rate $\gamma$ as shown in Fig.~\ref{Micromaser}(b).

In a single-atom micromaser, the interaction time between the atoms and the cavity field is typically fixed by the time it takes an atom to pass through the cavity. By contrast, here it fluctuates because of the random tunneling events. More precisely, the interaction time $\tau$ is Poisson distributed as $P(\tau) = \gamma e^{-\gamma \tau}$. Moreover, if electrons are evacuated much faster than they are injected, $\gamma \gg \Gamma$, we can employ a time-coarse-graining description of the time-evolution of the microwave cavity during the passage of a large number of electrons through the quantum dots. We then find that the dynamics of the cavity is given by the Lindblad equation (see Appendix~\ref{Appendix A})
\begin{eqnarray}
\nonumber
&\frac{d}{dt}\hat\rho  &= \mathcal{L}\hat \rho = -\frac{i}{\hbar}[\hat{H}_S,\hat \rho]+\kappa\left[(\nu+1)\mathcal D(\hat a)\hat\rho+\nu \mathcal D(\hat a^\dagger)\hat\rho\right]\\
 &&+\Gamma\left[\mathcal{D}\left(\sqrt{1-r[\hat a \hat a ^\dagger]}\right)\!\hat \rho+\mathcal{D}\left(\hat a^\dagger \frac{\sqrt{r[\hat a \hat a^\dagger]}}{\sqrt{\hat a \hat a^\dagger }}\right)\!\hat \rho \right]\!,
\label{Master equation}
\end{eqnarray}
where $\mathcal L$ is the Liouvillian, and $\hat \rho$ is the density matrix of the cavity. We have also introduced the dissipator 
\begin{equation}
\mathcal D(\hat a)\hat \rho \equiv \hat a \hat \rho \hat a ^\dagger-\frac{1}{2}\{\hat a^\dagger \hat a,\hat \rho\},
\end{equation}
where $\hat a$ is an annihilation operator that destroys photons in the cavity. In the Lindblad equation, the first term involving a commutator describes the coherent time-evolution given by the cavity Hamiltonian 
\begin{equation}
\hat H_S=\hbar \omega_0 (\hat a^\dagger \hat a+1/2).
\end{equation}
The second term describes the coupling to the thermal bath with strength $\kappa$ and equilibrium occupation number $\nu$ at temperature $T$. Finally, the third term is the average interaction between an electron and the cavity photons. We have also defined the operator
\begin{equation}
r[\hat a \hat a^\dagger] \equiv \frac{2\hat a \hat a^\dagger \varphi^2}{1+4\hat a \hat a^\dagger \varphi^2},
\label{r operator}
\end{equation}
which describes the probability that an electron transfers a photon to the cavity with $\varphi \equiv g/\gamma$ playing the role of the accumulated Rabi angle in a single-atom micromaser~\cite{Englert}.

\section{Full counting statistics}
\label{sec:fcs}

To understand the dynamic phase behavior of the micromaser, we show in Fig.~\ref{Phase diagrams}(a)-(c) Monte Carlo simulations of the number of electrons that have reached the right drain electrode in a given time span. Importantly, the number of transferred electrons equals the net number of photons emitted from the quantum dots to the cavity. As seen in the figure, this number depends strongly on the pump parameter $\alpha \equiv \varphi \sqrt{2\Gamma/\kappa}$, which quantifies the interaction strength between the electrons in the quantum dots and the photons in the microwave cavity. In the weak-pumping regime, $\alpha\ll 1$, a small fraction of the electrons going through the quantum dots are accompanied by the emission of a photon, resulting in a small current in the right drain. However, as the pumping strength increases, the probability of photon emission is enhanced. In particular, around $\alpha \simeq 1$, a dramatic change occurs, and the drain current grows strongly as the coupling is increased.

To explore the statistical properties of the emission process, it it useful to consider the generating function,
\begin{equation}
    \mathcal{G}(s,t) = \sum_{m=0}^\infty P(m,t) s^m,
    \label{Generating function}
\end{equation}
where $P(m,t)$ is the probability distribution of the number of electrons $m$ that have been collected in the right drain electrode during the time span $[0,t]$. Here, we will consider the \emph{factorial} moments and cumulants of $m$ by defining the factorial moment generating function
\begin{equation}
    \mathcal{M}_F(s,t) = \mathcal{G}(1+s,t) = \sum_{m=0}^\infty P(m,t) (1+s)^m.
    \label{FMGF}
\end{equation}
The factorial moments are then obtained by repeated differentiation with respect to $s$ at $s=0$,
\begin{equation}
    \langle m^k \rangle_F(t) = \left.\partial^k_s \mathcal{M}_F(s,t)\right|_{s=0},
\end{equation}
and they are related to the ordinary moments as 
\begin{equation}
    \langle m^k \rangle_F =  \langle m(m-1)\cdots(m-k+1) \rangle.
    \label{eq:ord2fac}
\end{equation}
For instance, the first two factorial moments are $\langle m\rangle_F=\langle m\rangle$ and $\langle m^2\rangle_F=\langle m^2\rangle-\langle m\rangle$, and we note that the ordinary moments can be obtained by differentiating the generating function $\mathcal{G}(s,t)$ with $s$ replaced by $e^s$ at $s=0$. 

The factorial cumulants are similarly encoded in the factorial cumulant generating function,
\begin{equation}
    \mathcal{F}_F(s,t) =  \ln \mathcal{M}_F(s,t),
    \label{eq:FCGF}
\end{equation}
and follow by differentiation with respect to $s$ as
\begin{equation}
    \cumu{m^k}_F = \left.\partial^k_s \mathcal{F}_F(s,t)\right|_{s=0}.
\end{equation}
The factorial cumulants are related to the ordinary ones just as the moments in Eq.~(\ref{eq:ord2fac}). Moreover, while only the first two cumulants are nonzero for a Gaussian distribution, only the first factorial cumulant is nonzero for a Poisson distribution. As we will see in the following, the factorial moments and cumulants provide a convenient description of the single-electron micromaser. In particular, while ordinary cumulants are useful to describe continuous random variables, factorial cumulants are in some cases better suited to characterize discrete random variables, such as the number of counted electrons or photons~\cite{PhysRevB.83.075432,Kambly2013,Stegmann2015,Stegmann2016,Stegmann2017,Kleinherbers2018,Kleinherbers2021,besson2021,stegmann2021,PhysRevB.94.125433,10.21468/SciPostPhys.10.1.007,PhysRevB.104.125431}.

To obtain the factorial moment generating function from the Lindblad equation [Eq.~\eqref{Master equation}], we first note that the coherences decouple from the populations 
\begin{equation}
p_n = \langle n|\hat \rho|n\rangle.
\end{equation}
We can then derive a  set of coupled equations for the populations alone reading
\begin{equation}
    \frac{dp_n(t)}{dt} = \gamma^+_{n-1}p_{n-1}(t)+\gamma^-_{n+1}p_{n+1}(t)- \left(\gamma^-_n+\gamma^+_n\right) p_n(t)
    \label{Pauli master equation}
\end{equation}
where we have introduced the effective absorption rate $\gamma^+_n=\kappa \nu(n+1)+\Gamma r[n+1]$ and the effective emission rate $\gamma^-_n =  \kappa(\nu+1)n$ for a cavity with $n$ photons.

Next, we introduce the probability $p_n(m,t)$ of having $n$ photons in the cavity after $m$ electrons have been transferred to the right drain during the time span $[0,t]$. By summing over the number of cavity photons, $P(m,t) = \sum_n p_n(m,t)$, we then find the full counting statistics of the number of electrons that have reached the right drain. From Eq.~\eqref{Pauli master equation}, we see that the probabilities $p_n(m,t)$ evolve according to the coupled equations
\begin{eqnarray}
\nonumber
    \frac{dp_n(m,t)}{dt} = \kappa \nu n p_{n-1}(m,t)+\Gamma r[n] p_{n-1}(m-1,t)\\
    +\gamma^-_{n+1}p_{n+1}(m,t)- \left(\gamma^-_n+\gamma^+_n\right) p_n(m,t),\hspace{3mm}
\end{eqnarray}
since every net absorption of a photon from the quantum dots increases the number of transferred electrons by one. This set of coupled equations may be decoupled with respect to $m$ by using the transformation 
\begin{equation}
p_n(s,t) \equiv \sum_{m} p_n(m,t) (1+s)^m.
\end{equation}
We then obtain
\begin{eqnarray}
\nonumber
    \frac{dp_n(s,t)}{dt} = \gamma^+_{n-1}(s)p_{n-1}(s,t)+\gamma^-_{n+1}p_{n+1}(s,t)\\
    - \left(\gamma^-_n+\gamma^+_n\right) p_n(s,t),
    \label{Counting field resolved master equation}
\end{eqnarray}
which is identical to Eq.~\eqref{Pauli master equation}, except that the absorption rate has been replaced by its counting-variable dependent counterpart,  $\gamma^+_n(s)=\kappa \nu(n+1)+\Gamma r[n+1] (1+s)$. Finally, we perform another transformation, 
\begin{equation}
G(s,q,t) = \sum_{n=0}^\infty p_n(s,t) e^{qn},
\end{equation}
to decouple the equations with respect to $n$. We then arrive at the partial differential equation
\begin{equation}
\begin{split}
    \partial_t G(s,q,t) = \big(&\kappa\nu(e^q-1)(1+\partial_q)+\kappa(\nu+1)(e^{-q}-1)\partial_q\\
    &+\Gamma \left(e^q(1+s)-1\right)r[\partial_q] \big) G(s,q,t),
\end{split}
    \label{Equation for G}
\end{equation}
whose solution at $q=0$ yields the factorial moment generating function as
\begin{equation}
    \mathcal{M}_F(s,t) = G(s,0,t) = \sum_n p_n(s,t).
\end{equation}
Generally, it is hard to solve Eq.~\eqref{Equation for G} for $G(s,q,t)$. However, as we show in the next section, it can be solved in  certain limits, allowing us to obtain analytic expressions for the factorial moment generating function. It will also be the starting point for our analysis of nonequilibrium phase transitions in the single-electron micromaser.

\section{Dynamical phases}
\label{Dynamical phases}

Figure~\ref{Micromaser}(a) shows the average current and the noise in the right drain as a function of the pump parameter $\alpha$, and it is indicative of a nonequilibrium phase transition occurring around $\alpha\simeq1$, where the average current develops a finite value, and the noise diverges as the injection rate $\Gamma$ is increased. Here, the average current and the noise are given by the long-time averages, $I_R=\lim_{t\rightarrow\infty} \langle m\rangle_F/t$ and $S_R=\lim_{t\rightarrow\infty} \cumu{m^2}_F/t$, respectively. (Notice that we here have used the second factorial cumulant instead of the ordinary cumulant, but this makes no qualitative difference.) Below $\alpha=1$, the current is vanishing, while it takes on a finite value above $\alpha=1$, and we will refer to each of those two phases as the low-activity and high-activity phases. To better understand the transition between them, we first treat the counting variable~$s$ as a tunable parameter in the spirit of the $s$-ensemble formalism of nonequilibrium phase transitions~\cite{PhysRevLett.104.160601,PhysRevE.84.021115,Merolle2005,Garrahan2007,Hedges2009,Speck2012}. Specifically, we consider the $s$-dependent current $I_R(s)$, obtained by replacing the average $\langle m\rangle$ in the current $I_R$ by the $s$-dependent expression
\begin{equation}
\langle m \rangle(s) = \partial_s \mathcal{M}_F(s,t)=\sum_{m=0}^\infty m P(m,t) (1+s)^{m-1}.
\end{equation}
While the actual physical dynamics takes place at $s=0$, positive and negative values of $s$ artificially enhance the probabilities of observing large or small values of $m$ compared to the real dynamics. Thus, using $s$ as a biasing field, we can enhance the rare events in the system. Moreover, as we will see, the behavior of the system away from $s=0$ in fact has measurable consequences. 

In Fig.~\ref{Phase diagrams}(e), we show a phase diagram of the current as a function of the pump parameter $\alpha$ and the counting field~$s$. Again, two distinct dynamical phases can be seen: the low-activity phase with a nearly vanishing current and the high-activity phase, where the current is on the order of the injection rate. In the following, we examine the properties of each phase before exploring the phase transition itself. Thus, below, we consider the full counting statistics for $\alpha \ll 1$ and for $\alpha \gg 1$, separately. We focus on the regime, where the injection rate is large compared to the cavity decay rate, $\Gamma\gg \kappa$, and the phase transition is pronounced.

\subsection{Low-activity phase}
For weak couplings to the cavity, $\alpha \ll 1$, the current in the right drain is close to zero, and the distribution $P(m,t)$ of transferred electrons is centered around $m=0$ as seen in Fig.~\ref{Phase diagrams}(d). In this regime, the interaction between the electrons and the photons is so weak that very few electrons are accompanied by the emission of a photon into the cavity. Consequently, the number of cavity photons is low, which in turn further suppresses the interactions with the electrons. We may also write Eq.~\eqref{Equation for G} as
\begin{eqnarray}
\nonumber
    \partial_t G(s,q,t) &=& \kappa\big(\nu(e^q-1)(1+\partial_q)+(\nu+1)(e^{-q}-1)\partial_q\\
    &+&\alpha^2 \left(e^q(1+s)-1\right)\partial_q \big) G(s,q,t),
\end{eqnarray}
having approximated the expression in Eq.~\eqref{r operator} as $r[\partial_q] \approx \alpha^2\partial_q \kappa/\Gamma$. Following Ref.~\onlinecite{PhysRevB.99.085418}, this partial differential equation can be solved using the method of characteristics. The factorial moment generating function then becomes
\begin{equation}
 \mathcal{M}_F(s,t) = \frac{2\xi e^{\tilde \kappa t/2}}{2\xi \cosh\left[\frac{\xi\tilde \kappa t}{2} \right]+(1+\xi^2)\sinh\left[\frac{\xi \tilde \kappa t}{2} \right]},
\label{FMGF linear regime}
\end{equation}
with $\tilde \kappa = (1-\alpha^2)\kappa$ and
\begin{equation}
    \xi = \sqrt{1-4\frac{\alpha^2}{\nu+\alpha^2}\langle n\rangle(1+\langle n\rangle)s},
    \label{xi expression}
\end{equation}
where the average number cavity photons reads
\begin{equation}
\langle n\rangle = \frac{\nu+\alpha^2}{1 -\alpha^2} \approx \nu+\alpha^2(1+\nu),
\label{Average photon number in cavity}
\end{equation}
recalling that $\nu$ is the equilibrium cavity occupation.

The overall structure of the factorial moment generating function determines the finite-time behavior of the current and resembles the emission statistics of photons from a thermal microwave cavity~\cite{PhysRevB.99.085418}, but with the time rescaled by the factor $1-\alpha^2$. Furthermore, we note that $\xi$ describes the long-time behavior as clearly visible by the scaled factorial cumulant generating function
\begin{equation}
 \Theta_F(s) = \lim_{t\rightarrow \infty} \frac{1}{t}\ln \mathcal{M}_F(s,t) = \frac{\kappa}{2} \left(1-\alpha^2\right)\left(1-\xi\right),
\end{equation}
which is consistent with the findings of Refs.~\onlinecite{Padurariu:2012,Arndt:2019,Arndt:2021}.
Differentiating Eq.~\eqref{FMGF linear regime} with respect to $s$, we obtain the average number of electrons transferred to the right drain
\begin{eqnarray}
    \langle m \rangle_F(t) &=& \kappa\alpha^2(1+\langle n\rangle)t,
    \label{Mean value for small alpha}
\end{eqnarray}
which indeed is small for $\alpha \ll 1$. Here, we see that the temperature of the thermal bath also plays an important role for the number of transferred electrons. At low temperatures, the average photon number in the cavity is small according to Eq.~\eqref{Average photon number in cavity}, and it increases with temperature. For low temperatures, $\nu \ll 1$, we may expand $\xi \approx 1-2\frac{\alpha^2}{\nu+\alpha^2}\langle n\rangle (1+\langle n \rangle)s$ around $s\simeq 0$ in Eq.~\eqref{xi expression}, and  the factorial cumulant generating function then becomes
\begin{equation}
     \Theta_F(s) \approx \langle I\rangle s,
\end{equation}
corresponding to the Poisson distribution 
\begin{equation}
    P(m,t) = \frac{\langle m\rangle^me^{-\langle m\rangle}}{m!}.
    \label{Low activity prob}
\end{equation}
with the average value given by Eq.~\eqref{Mean value for small alpha}, $\langle m\rangle=\langle m \rangle_F(t)$. As shown in the inset of Fig.~\ref{Phase diagrams}(d), this distribution captures the full counting statistics for $\alpha = 0.2$ and $\nu= 0.1$.

\subsection{High-activity phase}
In the opposite limit, $\alpha \gg 1$, where the interaction between the electrons and the photons is strong, the drain current is given by the injection rate as seen in Fig.~\ref{Phase diagrams}(e). We may then approximate the expression in Eq.~\eqref{r operator} by $r[\partial_q]\approx 1/2$. From Eq.~\eqref{Equation for G}, we then obtain 
\begin{eqnarray}
\nonumber
    \partial_t G(s,q,t) &=& \big(\kappa\nu(e^q-1)(1+\partial_q)+\kappa(\nu+1)(e^{-q}-1)\partial_q\\
    &&+\Gamma \left(e^q(1+s)-1\right)/2 \big) G(s,q,t),
\end{eqnarray}
which can directly be solved for $\mathcal{M}_F(s,t) = G(s,0,t)$ as
\begin{equation}
    \mathcal{M}_F(s,t) = e^{\Gamma ts/2}
\end{equation}
corresponding to a Poisson distribution as in Eq.~(\ref{Low activity prob}) with the average given by the first factorial moment,
\begin{equation}
    \langle m\rangle_F(t) = \Gamma t/2.
\end{equation}
We see that the average current equals half of the injection rate, since, on average, half of the electrons are accompanied by the emission of a photon, while the other half leaves the quantum dots via the left drain with no photon emission.

\begin{figure*}[t]
    \centering
    \includegraphics[width=0.95\textwidth]{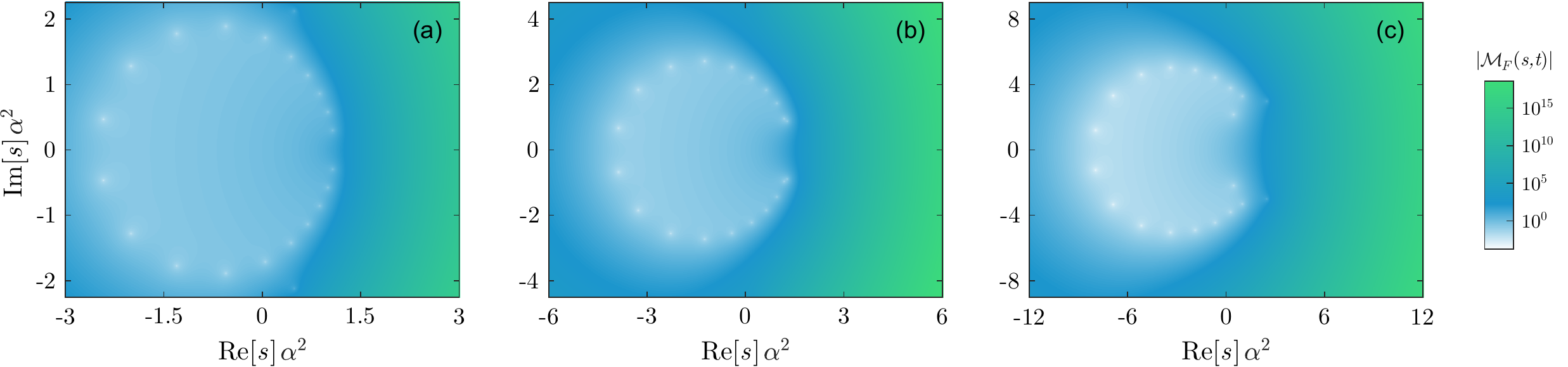}
    \captionsetup{justification=justified,singlelinecheck=false}
    \caption{Zeros of the factorial moment generating function. We show the absolute value of the factorial generating function in the complex plane of $s$ for (a) $\alpha = 0.1$, (b) $\alpha=0.5$ and (c) $\alpha = 1$ with $N=20$ cavity states and $\nu=0.1$, $\Gamma/\kappa=10$ and $t\kappa = 1$. The factorial moment generating function is plotted on a logarithmic scale, so that the zeros appear as white points.}
    \label{Fig3}
\end{figure*}

For long times, $\Gamma t\gg 1$, the Poisson distribution is well approximated by the Gaussian distribution
\begin{equation}
    P(m,t) = \frac{1}{\sqrt{2\pi\langle m\rangle}}\exp\left[-\frac{(m-\langle m\rangle)^2}{2\langle m\rangle} \right],
    \label{High activity prob}
\end{equation}
with the variance equal to the average. We expect a similar distribution will apply also for lower values of $\alpha$, however, with a modified average value related to the occupation of the cavity. We determine the occupation of the cavity  from the balance equation
\begin{equation}
    \Gamma r[1+\langle n\rangle]+\kappa \nu(1+\langle n\rangle) = \kappa (\nu+1)\langle n\rangle,
\end{equation}
having assumed that the cavity distribution is so narrow that we can use  the approximation $\langle r[1+n]\rangle \approx r[1+\langle n\rangle]$. We then find the average cavity photon number as
\begin{equation}
\begin{split}
    \langle n \rangle &= \frac{1}{8\varphi^2}\bigg(\alpha^2-1+4(\nu-1)\varphi^2\\
    +&\sqrt{1-2\alpha^2+8(1+\nu)\varphi^2+[\alpha^2+4(1+\nu)\varphi^2]^2}\bigg),
    \end{split}
\end{equation}
and the average number of transferred electrons becomes
\begin{equation}
    \langle m \rangle = \kappa (\langle n \rangle -\nu)t.
    \label{Average value big alpha}
\end{equation}
For $\alpha \gg 1$, we find $\langle n \rangle = \nu+\alpha^2/4\varphi^2 =\nu+ \Gamma/2\kappa$ and thus recover the expression $\langle m \rangle \approx \Gamma t/2$. Furthermore, as shown in Fig.~\ref{Phase diagrams}(c), the Gaussian distribution in Eq.~\eqref{High activity prob} with the average value set by Eq.~\eqref{Average value big alpha} is a good approximation of the full counting statistics already for $\alpha \gtrsim 3$.

\section{Nonequilibrium phase transition}
\label{Non-equilibrium phase transition}

Having characterized each of the two dynamical phases, we now go on to investigate the transition between them. To this end, we make use of Lee-Yang theory~\cite{Yang:1952,Lee:1952,Blythe2003,Bena:2005,Biskup2000,PhysRevLett.109.185701,PhysRevLett.114.010601} by considering the zeros of the factorial moment generating function~\cite{PhysRevB.83.075432,Kambly2013,Stegmann2015,Stegmann2016,Stegmann2017,Kleinherbers2018,Kleinherbers2021,besson2021,stegmann2021,PhysRevB.94.125433,10.21468/SciPostPhys.10.1.007,PhysRevB.104.125431},  which for our purposes plays the role of the partition function in equilibrium statistical mechanics~\cite{Flindt2013,Hickey2013,Hickey2014,Deger:2018,Deger:2019,Deger:2020,Deger:2020b,Peotta2021,Kist2021}. However, in contrast to thermal phase transitions, we are not considering transitions between different equilibrium phases such as spin lattices with a vanishing or a finite average magnetization. Instead, we consider transitions between different dynamical phases given by quantum jump trajectories that either have a low or a high dynamical activity~\cite{Merolle2005,Garrahan2007,Hedges2009,Speck2012}, here given by the average current. In this context, each quantum jump trajectory corresponds to a microstate of an equilibrium system. Moreover, the thermodynamic limit is reached as the number of states is increased combined with long observation times $t$, for which the quantum jump trajectories become very long.

In the original theory by Lee and Yang, equilibrium phase transitions are described by the zeros of the partition function in the complex plane of the control parameter, which move towards the critical value on the real axis as the thermodynamic limit is approached~\cite{Yang:1952,Lee:1952,Blythe2003,Bena:2005,Biskup2000,PhysRevLett.109.185701,PhysRevLett.114.010601}.  In a similar spirit, we will consider the zeros of the factorial moment generating function in the complex plane of the counting variable $s$. If the system exhibits a phase transition, the zeros will reach $s=0$ in the thermodynamic limit. As a result, the factorial cumulant generating function develops a singularity at $s=0$, just as how a singularity in the free energy signals an equilibrium phase transition.  Importantly, as we will see, we can determine the zeros from the factorial moments and cumulants of the drain current, which are measurable quantities.

\begin{figure*}
    \centering
    \includegraphics[width=0.95\textwidth]{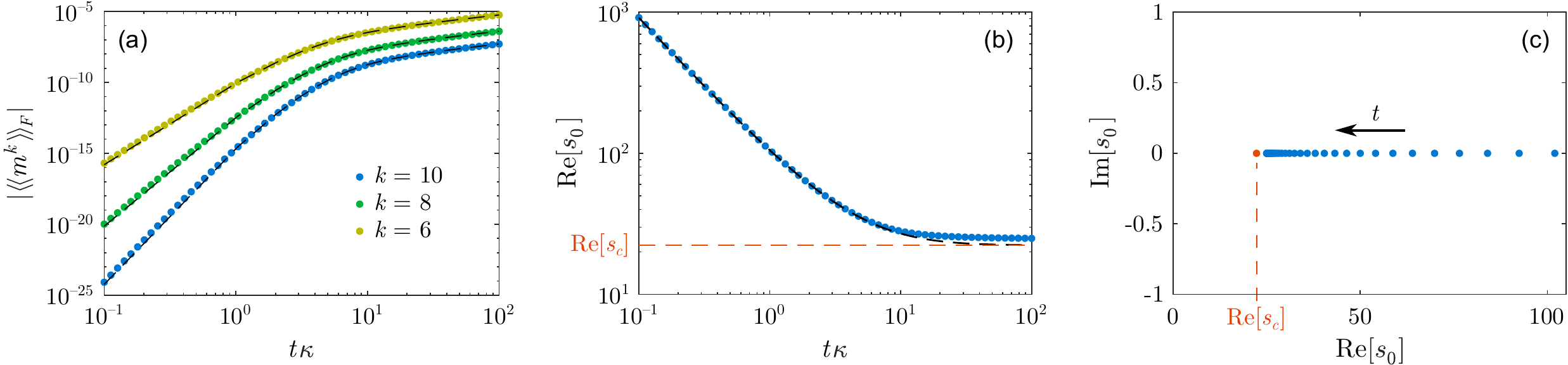}
    \captionsetup{justification=justified,singlelinecheck=false}
    \caption{Factorial cumulants and determination of poles. (a) Higher-order factorial cumulants $\langle \!\langle m^k \rangle\!\rangle$ of the drain current as functions of $t$ for $\alpha = 0.1$, $\nu=0.1$ and $\Gamma/\kappa = 250$. (b) The leading pole (closest to $s=0$) of the factorial moment generating function as a function of $t$, extracted from the factorial cumulants up to order $k=10$. (c) The leading pole for different times (blue) and the analytic convergence point in the infinite-time limit (red) in the complex plane of $s$. For the numerical calculations, $N=135$ cavity states were used. For comparison, the dashed lines in all three panels indicate the analytic results obtained from the factorial moment generating function in Eq.~\eqref{FMGF linear regime}.}
    \label{Cumulants and poles}
\end{figure*}

\subsection{Lee-Yang theory}
The starting point of our Lee-Yang theory is the factorial moment generating function in Eq.~\eqref{FMGF}. Similar to how the partition function encodes information about the fluctuations in an equilibrium system, the factorial moment generating function characterizes the fluctuations of the drain current. The partition function for a finite-size system is an entire function, and we similarly expect that the factorial moment generating function at finite times and system sizes will have no singularities in the complex plane of the counting variable. We may then expand it in terms of its the complex zeros $s_j$  as~\cite{Arfken2012}
\begin{equation}
    \mathcal{M}_F(s,t) = e^{c s} \prod \limits_j \left(1-s/s_j\right),
    \label{Factorization of MGF}
\end{equation}
where $c$ is a constant. The factorial cumulant generating function then becomes
\begin{equation}
    \mathcal{F}_F(s,t) = c s + \sum \limits_j \ln\left(1-s/s_j\right),
    \label{FMGF_LY}
\end{equation}
and the factorial cumulants read 
\begin{equation}
\langle \!\langle m^k\rangle\!\rangle_F =(k-1)!\sum_j \frac{e^{-ik\arg[s_j]}}{|s_j|^k},\quad k>1,
\label{Cumulant in terms of zeros and poles}
\end{equation}
having used polar coordinates, $s_j = |s_j|e^{i\arg[s_j]}$. From this expression, we see that the contribution of a zero to the factorial cumulants is suppressed with the distance $|s_j|$ to $s=0$ to the power of the cumulant order~$k$. Thus, for large $k\gg1$, the factorial cumulants are dominated by the zeros that are closest to $s=0$. Moreover, since the factorial moment generating function is real for real values of $s$, the zeros come in complex conjugate pairs. Using these properties, we may invert the relation in Eq.~\eqref{Cumulant in terms of zeros and poles} to find the real and absolute part of the closest conjugate pair of zeros, $s_0$ and $s_0^*$~\cite{Deger:2018,Deger:2019,Deger:2020,Deger:2020b},
\begin{widetext}
\begin{equation}
		\mathrm{Re}\left[s_{0}\right]\simeq \frac{k\left(k+1\right) \cumu{m^k}_F \cumu{m^{k+1}}_F -k\left(k-1\right) \cumu{m^{k-1}}_F   \cumu{m^{k+2}}_F }{2\left[ (k+1) \cumu{m^{k+1}}_F^2- \ k \cumu{m^k}_F  \cumu{m^{k+2}}_F\right]},  \quad k\gg 1,
		\label{eq:CumulantFormula1}
\end{equation}
and	
\begin{equation}
		\left|s_{0}\right|^2\simeq\frac{k^2 \left(k+1\right) \cumu{m^k}_F^2 - k\left(k^2-1\right) \cumu{m^{k-1}}_F  \cumu{m^{k+1}}_F}{(k+1) \cumu{m^{k+1}}_F^2- \ k \cumu{m^k}_F  \cumu{m^{k+2}}_F},  \quad k\gg 1.
		\label{eq:CumulantFormula2}
\end{equation}
\end{widetext}

These equations allow us to determine the dominant pair of zeros directly from four factorial cumulants of consecutive orders, and these are experimentally measurable quantities. Below, we evaluate the factorial cumulants using the numerical scheme described in Appendix \ref{Appendix B}.

\subsection{Poles in the generating function}
Before exploring the nonequilibrium phase transition around $\alpha=1$, we first consider the low-activity phase in the weak coupling regime, $\alpha\ll 1$, where we can compare our results with the analytic expressions in Sec.~\ref{Dynamical phases}. Considering the factorial moment generating function in Eq.~\eqref{FMGF linear regime}, we see that it in fact has no zeros, but only poles. As we will now explain, this is due to the infinite number of states of the microwave cavity that are all included in the derivation of Eq.~\eqref{FMGF linear regime}. As recently pointed out in the context of the Rabi model, quantum phase transitions can occur even for a two-level system coupled to a quantum harmonic oscillator~\cite{Hwang2015,Hwang2018}, and here the infinite number of oscillator states is crucial.

To understand how poles can emerge, we show in Fig.~\ref{Fig3}(a) the factorial moment generating function in the complex plane of $s$ for a finite number of photon states. The zeros are arranged in a circular shape, and the number of zeros increases as more photon states are included. This behavior is similar to a geometric series, $g_N(s) = \sum_{n=0}^{N} s^n$, with $N$ zeros on the unit circle, which eventually become a pole at $s=1$ for $N\rightarrow\infty$, since $g_\infty(s)=1/(1-s)$ for $|s|<1$. Similarly, in Eq.~\eqref{FMGF linear regime}, the zeros of the factorial moment generating function have transformed into poles. Nevertheless, when calculating the factorial cumulants, we get the same expression as in Eq.~(\ref{Cumulant in terms of zeros and poles}), only with the opposite sign and the zeros replaced by the poles. Therefore, we can still use  Eqs.~\eqref{eq:CumulantFormula1} and \eqref{eq:CumulantFormula2} to determine the poles that are closest to $s=0$. Moreover, if there is only a single dominant pole on the real axis, we can use the  simpler relation
\begin{equation}
    \mathrm{Re}[s_0] \simeq k \frac{\langle\!\langle m^k\rangle\!\rangle_F}{\langle \!\langle m^{k+1}\rangle\!\rangle_F}, \quad k\gg 1
    \label{Pole extraction}
\end{equation}
to determine it from two consecutive factorial cumulants. 

In Fig.~\ref{Cumulants and poles}(a), we show three higher-order factorial cumulants calculated as functions of time with a finite number of cavity states. From two consecutive cumulants, we then obtain the leading pole on the real axis using Eq.~(\ref{Pole extraction}) as shown in Fig.~\ref{Cumulants and poles}(b). With time the pole converges to the real axis, as illustrated in Fig.~\ref{Cumulants and poles}(c), with a value that is consistent with the convergence point
\begin{equation}
    s_c = \frac{(1-\alpha^2)^2}{4\alpha^2(1+\nu)}.
\end{equation} 
obtained from Eq.~\eqref{FMGF linear regime} in the long-time limit. As an additional check of our results, the dashed lines in Fig.~\ref{Cumulants and poles}(a)-(b), show analytic calculations based on the factorial moment generating function in Eq.~\eqref{FMGF linear regime}, which is valid when all cavity states are included. We note that for the regime considered here with $\alpha \ll 1$,  the convergence point is much larger than one for typical temperatures.  

\begin{figure*}[t]
    \centering
    \includegraphics[width=0.95\textwidth]{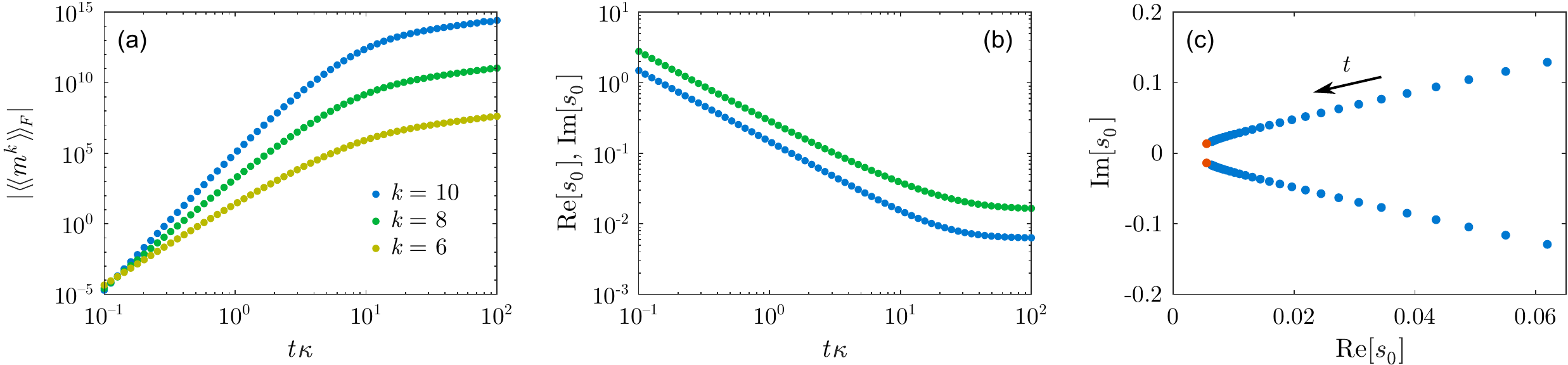}
    \captionsetup{justification=justified,singlelinecheck=false}
    \caption{Factorial cumulants and determination of Lee-Yang zeros. (a) Factorial cumulants $\langle \!\langle m^k \rangle\!\rangle$ of the drain current as functions of $t$ for $\alpha = 1$, $\nu=0.1$ and $\Gamma/\kappa = 250$. (b) The real (blue) and imaginary (green) parts of the leading zeros (closest to $s=0$) of the factorial moment generating function as a function of $t$, extracted from the cumulants up to tenth order. (c) The leading zeros for different times (blue) and the estimated convergence point (red) shown in the complex plane of $s$. For the numerical calculations, $N=135$ cavity states were used.}
    \label{Cumulants and zeros}
\end{figure*}

\subsection{Transition between the dynamical phases}
As shown in Fig.~\ref{Fig3}, the singular points of the factorial moment generating function move in the complex plane as $\alpha$ is increased. For moderate values, $\alpha \simeq 0.5$, there is a gap opening up in the circular structure of the zeros, which destroys the simple pole structure for $\alpha \ll 1$. As we further increase $\alpha$, and it gets closer to one, individual zeros start to move towards $s=0$. As a result, the cumulants are dominated by a conjugate pair of leading zeros instead of a pole. Thus, we employ Eqs.~\eqref{eq:CumulantFormula1} and \eqref{eq:CumulantFormula2} to determine the zeros from the factorial cumulants. 

\begin{figure*}[t]
    \centering
    \includegraphics[width=0.95\textwidth]{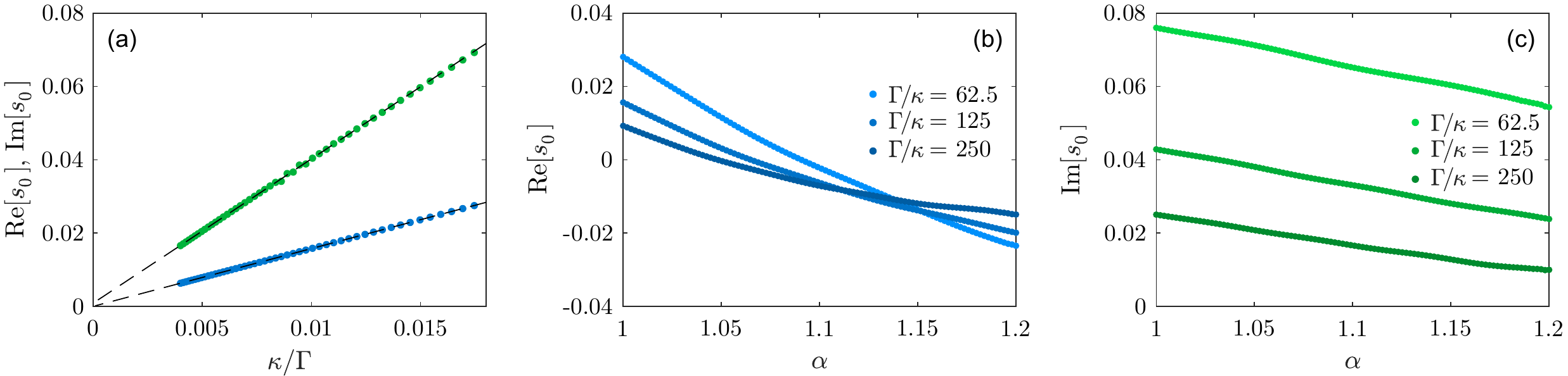}
    \captionsetup{justification=justified,singlelinecheck=false}
    \caption{Extrapolation of the Lee-Yang zeros. (a) The real (blue) and imaginary (green) parts of the convergence points as a function of $\kappa/\Gamma$ with $\alpha = 1$ and $\nu=0.1$. (b,c) The real (blue) and imaginary (green) parts of the convergence points as a function of $\alpha$ with $\Gamma/\kappa = 250$ and $\nu=0.1$. The real part for $\Gamma/\kappa = 250$ is indicated in the phase diagram in Fig.~\ref{Phase diagrams}(e).}
    \label{Zeros as a function of Gamma}
\end{figure*}

In Fig.~\ref{Cumulants and zeros}, we show the Lee-Yang zeros extracted for $\alpha = 1$. The factorial cumulants that we use are shown in Fig.~\ref{Cumulants and zeros}(a), while the extraction of the zeros is illustrated in Fig.~\ref{Cumulants and zeros}(b,c). As clearly seen, the zeros move with time towards the convergence points close to $s=0$ indicated with red circles. Performing this extraction for different values of  $\alpha$ close to $\alpha = 1$, we obtain the white points in Fig.~\ref{Phase diagrams}(e), which correspond to the real-part of the convergence points for different values of $\alpha$. These results, however, were obtained for a finite value of the injection rate $\Gamma$, for which the phase transition is not yet sharp. Thus, in Fig.~\ref{Zeros as a function of Gamma}(a), we show the real and imaginary parts of the convergence points as functions of $\kappa/\Gamma$ and in the limit, where this ratio goes to zero, the Lee-Yang zeros approach $s=0$, signalling that the system exhibits a nonequilibrium phase transition for $\alpha=1$. A further analysis is carried out in Figs.~\ref{Zeros as a function of Gamma}(b,c), where we examine the convergence points for values of $\alpha$ that are larger than one. Also in that case, the real and imaginary parts of the convergence points approach zero as the coupling $\Gamma$ is increased, and we conclude that the system exhibits a nonequilibrium phase transition along the dashed line in Fig.~\ref{Phase diagrams}(e). Thus, from the full counting statistics of tunneling events into the right drain we can predict the nonequilibrium phase transition taking place in the thermodynamic limit of long times. In contrast to the experiment of Ref.~\onlinecite{Brandner2017}, which involved only a few electronic states, the presence of the cavity makes it possible to reach the thermodynamic limit of large system sizes. Moreover, as shown in Fig.~\ref{Cumulants and zeros}, the results obtained at finite times can be extrapolated to the limit of long observation times, where the phase transition occurs.

\section{Experimental perspectives}
\label{sec:expper}

Finally, we comment on the perspectives of observing our predictions in an experiment using realistic system parameters. The frequency of the microwave cavity can be around $\omega_0/2\pi\simeq 10$~GHz, which also sets the energy difference between the excited state and the ground state of the double dots, if they are on resonance with the cavity. Furthermore, the tunnel coupling between the quantum dots can be around $T_0/\hbar\simeq 1$~GHz, which is much smaller than the energy splitting. (A too strong coupling would lead to delocalized eigenstates with a finite probability that an electron from the source lead is transferred into the ground state of the double dot system instead of the excited state.) With a bare coupling of $g_0/2\pi \simeq 1$~GHz~\cite{PhysRevA.69.042302,Bergenfeldt:2013}, the effective coupling between the quantum dots and the cavity photons becomes $g=g_0T_0/(\hbar \omega_0) \simeq 100$~MHz. It is also realistic to take $\gamma = 1$~GHz for the escape rate to the drain electrodes, $\Gamma =100$~MHz for the injection rate from the source electrode, and $\kappa = 1$~MHz for the cavity decay rate. In combination, we then obtain $\alpha = (g/\gamma)\sqrt{2\Gamma/\kappa}\simeq 1$, which is close to the critical value of $\alpha =1$. Furthermore, since $\Gamma/\kappa \gg 1$, the nonequilibrium phase transition should be rather sharp and visible. Altogether, these considerations show that our predictions should be within experimental reach using current technology.

\section{Conclusions and outlook}
\label{sec:conclusions}

We have shown that a nonequilibrium phase transitions in a single-electron micromaser can be predicted from the zeros of the factorial moment generating function, which can be determined by measuring the fluctuations of the drain currents at short times. To this end, we have made use of a cumulant method that is based on a extension of the classical Lee-Yang theory of equilibrium phase transitions to nonequilibrium settings. The phase transition can occur because of the coupling to a photon cavity, whose large state space allows us to reach the thermodynamic limit of large system sizes. 

Our work paves the way for systematic investigations of a broad range of physical systems that display nonequilibrium phase transitions. For instance, we expect that our method can be used to explore the superradiant phase transition in the Dicke model~\cite{Lambert:2004,Lambert:2005,Kirton:2019} and the quantum phase transition predicted for the Rabi model~\cite{Hwang2015,Hwang2018}. Moreover, since our approach is based on accurate measurements of fluctuating observables, we believe that it can also be realized in future experiments to observe a nonequilibrium phase transition in a solid-state device.

\acknowledgements
We thank L.~Otten and F.~Hassler for useful comments on the manuscript. F.~B.~acknowledges support from the European Union’s Horizon 2020 research and innovation programme under the Marie Skłodowska--Curie grant agreement No.~892956. The work was supported by Academy of Finland through the Finnish Centre of Excellence in Quantum Technology (project numbers 312057 and 312299) and grants number 308515 and 331737.

\appendix
\section*{Appendices}

\section{Derivation of the Lindblad equation}
\label{Appendix A}
Here we derive the Lindblad equation in Eq.~\eqref{Master equation} of the main text along the same lines as the derivation of the Lindblad equation for a single-atom micromaser, see, e.g.,~Ref.~\onlinecite{Englert}. We first consider the time-evolution of the closed microwave cavity-quantum dot system, with one electron being trapped inside the double dot system and the thermal bath decoupled from the cavity. Based on this, we infer the average effect on the dynamics of the microwave cavity induced by an electron inside the quantum dot system. Using time-coarse-graining, we then derive a Lindblad equation that governs the time-evolution of the reduced microwave-cavity system, with the effect of the thermal bath and the Poissonian flow of electrons through the quantum dot system included.

Our starting point is the Hamiltonian~\cite{PhysRevA.69.042302,Bergenfeldt:2013}
\begin{equation}
\hat H' = \hbar \omega_0 (\hat a^\dagger \hat a+1/2) + \frac{ \varepsilon}{2}\hat \sigma_z + \hbar g_0 \left(\hat a +\hat a^\dagger\right)\left(\frac{\varepsilon}{2\epsilon}\hat \sigma_z + \frac{ T_0}{\epsilon} \hat \sigma_x\right),
\end{equation}
describing a double quantum dot system, occupied by a single electron, capacitively coupled to a single-mode microwave cavity. The first two terms in the Hamiltonian describe the energy of the microwave cavity and the quantum dot system, respectively, while the third term describes the interaction energy due to the capacitive coupling between the photon field of the cavity and the charge of the quantum dot system. Here, $\hat a^\dagger$ creates a photon of angular frequency $\omega_0$ in the microwave cavity, $\hat \sigma_x$, $\hat \sigma_y$ and $\hat \sigma_z$ denote the Pauli matrices acting on the dot system, and $g_0$ is the coupling strength between the dot system and the microwave cavity. Furthermore, $\varepsilon$ denotes the potential-energy difference between the dots when uncoupled, and $\epsilon = \sqrt{\varepsilon^2+4T_0^2}$ the energy difference between the two eigenstates of the coupled dots. We will in the following consider $T_0\ll \varepsilon$, such that $\epsilon \approx \varepsilon$ and the eigenstates are approximately equal to the localized states of the dots.

Using the rotating-wave approximation in the interaction picture, i.e., neglecting the fast oscillating terms with frequency $\omega_0+\varepsilon/\hbar$, and then going back to the Schrödinger picture, we obtain the Jaynes-Cummings Hamiltonian, reading (up to an irrelevant constant)
\begin{equation}
\hat H = \hbar \omega_0 \hat a^\dagger \hat a + \varepsilon \hat \sigma^\dagger \hat \sigma + \hbar g \left(\hat a^\dagger \hat \sigma+\hat \sigma^\dagger \hat a\right),
\end{equation}
where $\hat \sigma^\dagger = \left(\hat \sigma_x +i \hat \sigma_y\right)/2$ and $\hat \sigma = \left(\hat \sigma_x -i \hat \sigma_y\right)/2$ are the excitation and de-excitation operators for the quantum dots, and $g= g_0 T/\varepsilon$ is the effective coupling strength for the dot-cavity interaction.

For the sake of simplicity, we now assume that the quantum dot system is in resonance with the microwave cavity, $ \varepsilon =\hbar \omega_0$. We may then express the Hamiltonian in terms of the transition operator $\hat \gamma = \hat a^\dagger \hat \sigma+\hat \sigma^\dagger \hat a$ as
\begin{equation}
\hat H = \hbar \omega_0 \hat \gamma^2 +\hbar g \hat \gamma.
\end{equation}
Since $[\hat H,\hat \gamma] = 0$, it is possible to find a common basis of eigenstates for $\hat H$ and $\hat \gamma$. These eigenstates are
\begin{equation}
|\gamma_n^\pm \rangle = \frac{1}{\sqrt{2}}\left(|L,n\rangle \pm |R,n+1\rangle  \right),
\end{equation}
where $|L\rangle$ and $|R\rangle$ are eigenstates of $\hat \sigma^\dagger \hat \sigma$ (with $\hat \sigma^\dagger \hat \sigma|L\rangle = |L\rangle$ and $\hat \sigma^\dagger \hat \sigma|R\rangle = 0$) and $|n\rangle$ is the $n$th eigenstate of $\hat a^\dagger \hat a$. Note that we are here making use of the assumption $T_0\ll \varepsilon$, where the eigenstates of the dot system are approximately equal to the localized states of the dots.

Since $\hat \gamma$ is Hermitian, the eigenstates form a complete orthogonal basis, and the corresponding eigenvalues are
\begin{equation}
\lambda_n^\pm = \pm \sqrt{n+1}.
\end{equation}
The time-evolution of these eigenstates in the interaction picture (with $\hat H_\text{int} = \hbar g \hat \gamma$) is
\begin{equation}
i\hbar \partial_t |\gamma_n^\pm (t) \rangle = \hat H_\text{int} |\gamma_n^\pm (t)\rangle = \hbar g \lambda_n^\pm |\gamma_n^\pm (t)\rangle,
\end{equation}
such that
\begin{equation}
|\gamma_n^\pm (t) \rangle = e^{-i g \lambda_n^\pm(t-t_0)}|\gamma_n^\pm (t_0)\rangle.
\end{equation}
\begin{widetext}
For $|L,n,t\rangle$, the time-evolution is
\begin{eqnarray}
\nonumber
|L,n,t\rangle &=& \frac{1}{\sqrt{2}}\left[|\gamma_n^+(t)\rangle+|\gamma_n^-(t)\rangle\right]=\frac{1}{\sqrt{2}}\Big[e^{-i g \lambda_n^+(t-t_0)}|\gamma_n^+ (t_0)\rangle +e^{-i g \lambda_n^-(t-t_0)}|\gamma_n^- (t_0)\rangle\Big]\\
&=& \cos\Big[g\sqrt{n+1}(t-t_0)\Big]|L,n,t_0\rangle -i \sin\Big[g \sqrt{n+1}(t-t_0)\Big] |R,n+1,t_0\rangle.\hspace{5mm}
\label{Time evolution of L state}
\end{eqnarray}
This expression shows how the microwave cavity-quantum dot system evolves with an electron in the quantum dots.

We now derive the average effect of an electron on the reduced density matrix of the microwave cavity. To this end, we consider the density matrix of the combined microwave cavity-quantum dot system right at the moment an electron has tunneled into the left quantum dot from the source lead. Using that $T_0 \ll \varepsilon$, the electron is initially in the excited state of the double dot system,
\begin{equation}
\hat \rho^\text{tot} = \sum_{m} \rho^\text{tot}_{m} |L,m ,0\rangle   \langle L,m ,0|,
\end{equation}
where we have assumed that there are no initial coherences between different photon number states, i.e., $\rho^\text{tot}_{m,n} =0$ for $m\neq n$. From Eq.~\eqref{Time evolution of L state}, we find that the state after an electron has interacted with the cavity for the time $\tau$ reads (in the interaction picture)
\begin{eqnarray}
\nonumber
\hat \rho^\text{tot}(\tau) &=& \sum_{m} \rho^\text{tot}_{m}\Big(\cos\left[g \sqrt{m+1}\tau\right]|L,m ,\tau\rangle-i \sin\left[g \sqrt{m+1}\tau\right] |R,m+1 ,\tau\rangle \Big) \\
&&\times  \Big(\cos\left[g \sqrt{m+1}\tau\right]\langle L,m ,\tau|+i \sin\left[g \sqrt{m+1}\tau\right] \langle R,m+1 ,\tau| \Big).
\end{eqnarray}
Tracing out the quantum dot system, we find the reduced density matrix of the microwave cavity
\begin{eqnarray}
\nonumber
\hat \rho(\tau) &=&\langle L |\hat \rho^\text{tot}(\tau)|L\rangle +\langle R |\hat \rho^\text{tot}(\tau)|R\rangle \\
&=& \sum_{m}\rho_{m} \bigg( \cos^2\left[g\sqrt{m+1}\tau \right] |m\rangle \langle m|+\sin^2\left[g \sqrt{m+1}\tau\right] |m+1\rangle \langle n+1|\bigg).
\end{eqnarray}
We note that the interaction with the quantum dot system does not produce any coherences, and hence the state remains diagonal in the photon number basis.

Next, we average over the distribution of times that electrons spend in the quantum dots, $P_\gamma(\tau) = \gamma e^{-\gamma\tau}$, to obtain the average change $\Delta \hat \rho$ of the microwave cavity state induced by the electrons. Using that
\begin{equation}
\nonumber
    \int_0^\infty d\tau \cos^2\left[g\sqrt{m+1}\tau \right] \gamma e^{-\gamma \tau} = 1-\frac{2(m+1) g^2}{\gamma^2+4(m+1)g^2} \,\,\mathrm{and}\,\,
    \int_0^\infty d\tau \sin^2\left[g\sqrt{m+1}\tau \right] \gamma e^{-\gamma \tau} = \frac{2(m+1) g^2}{\gamma^2+4(m+1)g^2},
\end{equation}
we obtain
\begin{eqnarray}
\nonumber
\Delta \hat \rho&=&\int_{0}^\infty d\tau \hat \rho(\tau)P_\gamma(\tau) -\hat \rho \\
\nonumber
&=& \sum_{m} \rho_{m} \Bigg[ \left(1-\frac{2(m+1) g^2}{\gamma^2+4(m+1)g^2}\right) |m\rangle  \langle m| +\frac{2(m+1) g^2}{\gamma^2+4(m+1)g^2}|m+1\rangle \langle m+1| \Bigg] -\hat \rho \\
&=&  \sqrt{1-r[\hat a \hat a^\dagger]} \hat\rho\sqrt{1-r[\hat a \hat a^\dagger]}+\hat a^\dagger \frac{\sqrt{r[\hat a \hat a^\dagger]}}{\sqrt{\hat a \hat a^\dagger}}\rho  \frac{\sqrt{r[\hat a \hat a^\dagger]}}{\sqrt{\hat a \hat a^\dagger}}\hat a -\hat \rho,
\end{eqnarray}
where we have defined the operator $r[\hat a \hat a^\dagger] \equiv \frac{2\hat a \hat a^\dagger \varphi^2}{1+4\hat a \hat a^\dagger \varphi^2}$ with $\varphi = g/\gamma$. Introducing the Lindblad dissipator $\mathcal{D}(\hat a) \hat \rho \equiv \hat a \hat \rho \hat a^\dagger -\frac{1}{2}\left\{ \hat a^\dagger \hat a,\hat \rho \right\}$, we may write the result as
\begin{eqnarray}
\Delta \hat \rho&=& \mathcal{D}\left(\sqrt{1-r[\hat a \hat a ^\dagger]}\right)\!\hat \rho+\mathcal{D}\left(\hat a^\dagger \frac{\sqrt{r[\hat a \hat a^\dagger]}}{\sqrt{\hat a \hat a^\dagger }}\right)\!\hat \rho.
\label{Change due to one electron}
\end{eqnarray}
\end{widetext}

Finally, to obtain the generalized master equation, we note that for $\gamma \gg \Gamma$, the electrons are injected into the quantum dots in an uncorrelated way. Then, by employing time-coarse-graining along the lines of Ref.~\onlinecite{Englert}, we directly obtain the generalized master equation as
\begin{equation}
\frac{d}{dt}\hat\rho = -\frac{i}{\hbar}[\hat{H}_S,\hat \rho]+\kappa\left[(\nu+1)\mathcal D(\hat a)\hat\rho+\nu \mathcal D(\hat a^\dagger)\hat\rho\right]+\Gamma \Delta \hat \rho,
\end{equation}
where we have also included a thermal bath~\cite{breuer2007theory}. Here, $\hat H_S = \hbar \omega_0 (\hat a^\dagger \hat a+1/2)$ is the Hamiltonian of the microwave cavity, $\kappa$ is the coupling strength between the cavity and the bath, and $\nu$ is the equilibrium occupation number of the cavity. The first term describes the coherent time-evolution of the cavity, the second one the interactions with the thermal bath, and the third term describes the interaction with the quantum dots as given by Eq.~\eqref{Change due to one electron}.

\section{Numerical scheme for calculating the factorial cumulants at finite times}
\label{Appendix B}

For the sake of completeness, we here describe the numerical scheme for calculating factorial cumulants at finite times. To this end, we employ the numerical approach developed in Ref.~\onlinecite{PhysRevB.83.075432}. Our starting point is the matrix representation of the master equation in Eq.~\eqref{Counting field resolved master equation}, which can be expressed as
\begin{equation}
    \partial_t |p(s,t)\rangle\!\rangle = \mathbf{L}(s) |p(s,t)\rangle\!\rangle,
    \label{Matrix representation of gen master equation}
\end{equation}
where the matrix $\mathbf{L}(s)$ contains counting-field dependent transition rates, and $|p(s,t)\rangle\!\rangle$ is the vector representation of the counting-field dependent occupation probabilities. Importantly, the factorial moments may be expressed as
\begin{equation}
    \langle m^k \rangle_F = \langle\!\langle \tilde 0|p^{(k)}(t)\rangle\!\rangle,
\end{equation}
where $|p^{(k)}(t)\rangle\!\rangle \equiv \partial_s^k |p(s,t)\rangle\!\rangle\big|_{s=0}$ and $\langle\!\langle \tilde 0| = [1,1,...,1]$ is the vector representation of the trace operator, consisting of a vector with $N$ elements of ones. The time-evolution of $|p^{(k)}(t)\rangle\!\rangle$ is obtained by differentiating Eq.~\eqref{Matrix representation of gen master equation} $k$ times with respect to $s$ and evaluating the result at $s=0$,
\begin{equation}
    \partial_t |p^{(k)}(t)\rangle\!\rangle = \sum_{j=0}^k \begin{pmatrix}k\\j\end{pmatrix} \mathbf{L}^{(j)} |p^{(k-j)}(t)\rangle\!\rangle, \qquad k\geq 0,
    \label{Time evolution of p^k(t)}
\end{equation}
where $\mathbf{L}^{(k)} \equiv \partial_s^k \mathbf{L}(s)\big|_{s=0}$. To calculate the first $k$ factorial moments, we solve these coupled equations for $\{|p^{(j)}(t)\rangle\!\rangle\}_{j=0}^k$. To this end, we formulate the coupled equations as one single matrix equation of the form
\begin{equation}
    \partial_t |P(t)\rangle\!\rangle = \mathbf{\tilde L} |P(t)\rangle\!\rangle,
    \label{Coupled equations for p^(k)}
\end{equation}
having introduced the vector $|P(t)\rangle\!\rangle \equiv [|p^{0}(t)\rangle\!\rangle, |p^{1}(t)\rangle\!\rangle,...,|p^{k}(t)\rangle\!\rangle]$ together with the enlarged matrix
\begin{equation}
    \mathbf{\tilde L} \equiv \begin{pmatrix}
    \mathbf{L}^{(0)} & 0 & 0 & 0 & 0 \\
    \mathbf{L}^{(1)} & \mathbf{L}^{(0)} & 0 & 0 & 0 \\
    \mathbf{L}^{(2)} & 2\mathbf{L}^{(1)} & \mathbf{L}^{(0)} & 0 & 0 \\
    \vdots & \vdots & & \ddots &  \\
    \mathbf{L}^{(k)} & & & k\mathbf{L}^{(1)} & \mathbf{L}^{(0)}\\
    \end{pmatrix}.
\end{equation}
We use the initial condition $|P(0)\rangle\!\rangle = [|0\rangle\!\rangle,0,...,0]$, where $|0\rangle\!\rangle$ is the stationary state given by $\mathbf{L}(0)|0\rangle\!\rangle = 0$ and find
\begin{equation}
    |P(t)\rangle\!\rangle = e^{\mathbf{\tilde L}t}|P(0)\rangle\!\rangle,
\end{equation}
 which allows us to easily compute $|p^{(k)}(t)\rangle\!\rangle$ and thus $\langle m^k\rangle_F$. From the factorial moments, we finally obtain the cumulants using the standard relation between moments and cumulants
\begin{equation}
    \langle\! \langle m^k \rangle\!\rangle_F = \langle m^k\rangle_F -\sum_{j=1}^{k-1} \begin{pmatrix}k-1\\j-1\end{pmatrix}\langle\!\langle m^j\rangle\!\rangle_F\langle m^{k-j}\rangle_F.
\end{equation}


%

\end{document}